\documentclass[letterpaper, 10 pt, conference]{ieeeconf}  

\IEEEoverridecommandlockouts                             
\usepackage{float}
\usepackage{verbatim}
\usepackage{paralist}
\usepackage{amsfonts}
\usepackage{graphics} 
\usepackage{epsfig} 
\usepackage{mathptmx} 
\usepackage{times} 
\usepackage{amsmath} 
\usepackage{amssymb}  
\usepackage{physics}
\usepackage{mdwmath}
\usepackage{mdwtab}
\usepackage{color}
\usepackage[hidelinks]{hyperref}
\usepackage{algpseudocode}
\usepackage{algorithm}
\usepackage{caption}
\usepackage{subcaption}
\usepackage{amsmath} 
\usepackage{bm}      

\usepackage{tikz}
\usetikzlibrary{shapes, arrows.meta, positioning}

\def\BibTeX{{\rm B\kern-.05em{\sc i\kern-.025em b}\kern-.08em
    T\kern-.1667em\lower.7ex\hbox{E}\kern-.125emX}}
\begin{document}

\makeatletter
\newcommand{\newlineauthors}{%
  \end{@IEEEauthorhalign}\hfill\mbox{}\par
  \mbox{}\hfill\begin{@IEEEauthorhalign}
}
\newcommand{\compactblockA}[1]{%
  {\small \begin{tabular}{@{}c@{}}#1\end{tabular}}%
}

\makeatother


\title{Quantum-Assisted Trainable-Embedding Physics-Informed Neural Networks for Parabolic PDEs}

\author{Ban Q. Tran, Nahid Binandeh Dehaghani, Rafal Wisniewski, Susan Mengel, and A. Pedro Aguiar
\thanks{Ban Q. Tran is with the Department of Computer Science, Texas Tech University, Lubbock, USA, and Department of Computing Fundamentals, FPT University, Hanoi, Vietnam.
        {\tt\small bantran@ttu.edu - bantq3@fe.edu.vn}}
\thanks{Nahid Binandeh Dehaghani is with the Department of Electronic Systems, Aalborg University, Aalborg, Denmark
        {\tt\small nahidbd@es.aau.dk}}%
\thanks{Rafal Wisniewski is with the Department of Electronic Systems, Aalborg University, Aalborg, Denmark. {\tt\small raf@es.aau.dk}}
\thanks{Susan Mengel is with the Department of Computer Science, Texas Tech University, Lubbock, USA. {\tt\small susan.mengel@ttu.edu}} 
\thanks{A. Pedro Aguiar is with SYSTEC-ARISE, Faculty of Engineering, University of Porto, Porto, Portugal. {\tt\small 	pedro.aguiar@fe.up.pt}}
}

\maketitle

\begin{abstract}
Physics-informed neural networks (PINNs) have emerged as a powerful framework for solving partial differential equations (PDEs) by embedding governing physical laws directly into the training objective. Recent advances in quantum machine learning have motivated hybrid quantum–classical extensions aimed at enhancing representational capacity while remaining compatible with near-term quantum hardware. In this work, we investigate trainable embedding strategies within quantum-assisted PINNs for solving parabolic PDEs, using one- and two-dimensional Heat equations as canonical benchmarks.
We introduce two quantum-assisted architectures that differ in their embedding components. In the first approach, a classical feed-forward neural network generates trainable feature maps for quantum data encoding (FNN-TE-QPINN). In the second, the embedding stage is realized entirely by a parameterized quantum circuit (QNN-TE-QPINN), yielding a fully quantum feature map.
Our findings emphasize the critical role of embedding design and support hybrid quantum–classical approaches for parabolic PDE modeling in the NISQ era.
\end{abstract}

\section{Introduction}
The integration of quantum computing into scientific machine learning has opened new directions for solving complex differential equations. In particular, hybrid quantum–classical strategies \cite{kyriienko2021solving} have gained attention as a practical pathway for leveraging parameterized quantum circuits within classical optimization frameworks. At the same time, physics-informed neural networks (PINNs) have established themselves as effective tools for approximating solutions of partial differential equations (PDEs) by incorporating governing laws directly into the loss function.

Recent developments have combined these two paradigms by embedding variational quantum circuits into physics-informed architectures \cite{dehaghani2025quantumqce2025, tran2026trainableembedding, dehaghani2024hybrid,berger2025trainable}. Among these approaches, trainable-embedding quantum PINNs (TE-QPINNs) utilize classical or quantum feature maps to encode input coordinates into quantum states prior to variational processing. This design enables nonlinear function approximation while remaining compatible with noisy intermediate-scale quantum (NISQ) devices.

Despite encouraging results for nonlinear and coupled systems, a systematic investigation of embedding strategies for parabolic PDEs remains limited. Parabolic equations, such as the Heat equation, model diffusion-driven dynamics and serve as fundamental benchmarks for time-dependent spatiotemporal learning methods.
In this study, we adapt a quantum-assisted trainable-embedding framework to solve one- and two-dimensional Heat equations. 

This work builds upon our recent work on hybrid QPINN framework for time-dependent parabolic PDEs \cite{dehaghani2025quantumqce2025}. In that study, classical neural networks were employed exclusively for input embedding, while variational quantum circuits performed nonlinear transformation. The present work extends that framework by systematically investigating embedding design within quantum-assisted PINNs. In particular, we introduce and compare hybrid classical embedding and fully quantum embedding strategies under identical variational circuit and training configurations, enabling a controlled evaluation of their impact on convergence, accuracy, and parameter efficiency.
A schematic overview of the proposed learning architecture is provided in Fig.~\ref{fig:architecture}. 
\begin{figure}[t]
    \centering
    \includegraphics[scale=0.35]{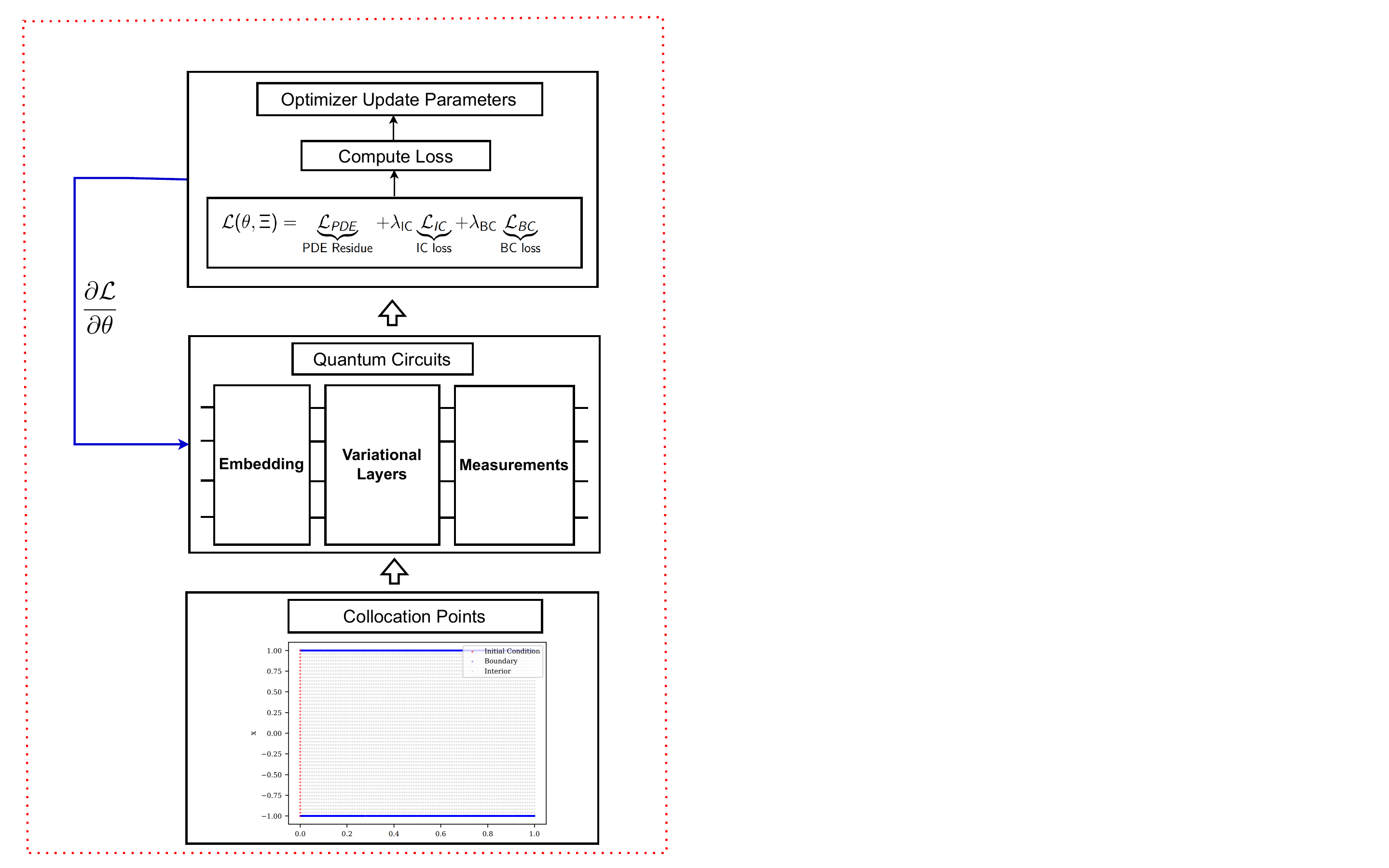}
    \caption{
    \small A hybrid classical quantum trainable embedding physics-informed neural networks architecture. Classical embedding maps collocation points into angle parameters for a parameterized quantum circuit. The expectation values are used to compute the residual loss, which is minimized using a classical optimizer.}
    \label{fig:architecture}
    \vspace{-0.5cm}
\end{figure}
The primary contributions of this work are as follows:
\begin{itemize}
    \item We investigate quantum-assisted physics-informed learning for parabolic PDEs and provide comprehensive numerical validation on one- and two-dimensional Heat equations.    
    \item We conduct a controlled comparison between hybrid classical embedding and fully quantum embedding within a shared variational quantum architecture.    
    \item We analyze convergence behavior, solution accuracy, and parameter efficiency under realistic NISQ-era simulation constraints.   
    \item We demonstrate that hybrid embedding strategies offer improved stability and predictive performance relative to both classical PINNs and purely quantum embedding approaches.    
    \item We provide architectural insights into the role of embedding mechanisms in quantum-enhanced PDE solvers.
\end{itemize}
The remainder of this paper is organized as follows. Section II presents the theoretical formulation of the parabolic PDE problem and introduces the proposed quantum-assisted learning framework, including the variational circuit structure, embedding strategies, observable design, and physics-informed optimization procedure. Section III provides comprehensive numerical experiments for one- and two-dimensional Heat equations, along with a detailed performance comparison between classical PINNs, hybrid FNN-TE-QPINNs, and fully quantum QNN-TE-QPINNs. Section IV concludes the paper with a summary of findings and discusses implications for embedding design in quantum-enhanced PDE solvers, as well as directions for future research.

\section{Theory and Methodology}

This section presents the mathematical formulation and the proposed quantum-assisted learning framework for solving parabolic partial differential equations. We first introduce the general residual formulation of the governing equation. We then describe the hybrid quantum–classical parametric representation of the solution, detail the embedding strategies under investigation, and finally discuss how architectural isolation enables a controlled evaluation of embedding design.

\subsection{Problem Formulation}

We consider a class of time-dependent parabolic partial differential equations posed on a spatial domain $\Omega \subset \mathbb{R}^d$ over a finite time horizon $t \in [0,T]$. The problem is defined through its residual representation:
\begin{align*}
\mathcal{R}_{\text{PDE}}(\mathbf{x},t) 
&:= \mathcal{D}\big(u(\mathbf{x},t);\beta\big) - q(\mathbf{x},t) = 0, 
\quad (\mathbf{x},t) \in \Omega \times (0,T], \\
\mathcal{R}_{\text{BC}}(\mathbf{x},t) 
&:= \mathcal{B}\big(u(\mathbf{x},t)\big) - b(\mathbf{x},t) = 0, 
\quad (\mathbf{x},t) \in \partial\Omega \times (0,T], \\
\mathcal{R}_{\text{IC}}(\mathbf{x}) 
&:= u(\mathbf{x},0) - u_0(\mathbf{x}) = 0, 
\quad \mathbf{x} \in \Omega.
\end{align*}
Here, $\mathbf{x}$ denotes spatial coordinates, $u(\mathbf{x},t)$ is the unknown field to be approximated, and $\mathcal{D}$ represents a differential operator involving temporal and spatial derivatives parameterized by physical constants $\beta$. The operators $\mathcal{B}$, $q$, $b$, and $u_0$ encode boundary, source, and initial conditions.
Our objective is to construct a parametric approximation $\tilde{u}(\mathbf{x},t)$ that minimizes these residuals across the spatiotemporal domain using a hybrid quantum–classical architecture.

\subsection{Quantum Parametric Representation}
The approximate solution is generated from a parameterized quantum circuit whose expectation value defines the scalar field prediction:

\[
\tilde{u}(\mathbf{x},t;{\theta}_{\mathrm{var}},{\theta}_{\mathrm{emb}})
=
\langle \psi(\mathbf{x},t) | \mathcal{O} | \psi(\mathbf{x},t) \rangle,
\]
where the quantum state is prepared as

\[
|\psi(\mathbf{x},t)\rangle 
=
U_{\text{var}}({\theta}_{\mathrm{var}}) 
\, U_{\text{enc}}(\mathbf{x},t;{\theta}_{\mathrm{emb}}) 
\, |0\rangle^{\otimes n}.
\]
Here, $U_{\text{enc}}$ denotes the input-dependent encoding unitary, $U_{\text{var}}$ is a trainable variational unitary acting on $n$ qubits, and $\mathcal{O}$ is a Hermitian observable whose expectation value defines the predicted solution.
The complete parameter set of the model is therefore given by 
$\Theta = ({\theta}_{\mathrm{var}}, {\theta}_{\mathrm{emb}})$,
where ${\theta}_{\mathrm{var}}$ and ${\theta}_{\mathrm{emb}}$ represent the variational and embedding parameters, respectively.

\subsection{Embedding Mechanisms}
The encoding stage determines how continuous spatiotemporal coordinates $(\mathbf{x},t)$ are mapped into quantum states prior to variational processing. Since the model output depends on the input exclusively through the encoding unitary, the structure of this mapping directly affects representational capacity, gradient propagation, and overall learning dynamics.
Let $\Gamma(\mathbf{x},t) = [\phi_1(\mathbf{x},t), \ldots, \phi_n(\mathbf{x},t)]^\top$ denote a vector of rotation angles applied to an $n$-qubit register. 
The encoding unitary is defined as
\[
U_{\text{enc}}(\mathbf{x},t;{\theta}_{\mathrm{emb}})
=
\prod_{k=1}^{n} R_y\big(\phi_k(\mathbf{x},t)\big),
\]
where $R_y(\cdot)$ denotes a single-qubit rotation about the $y$-axis. The parameters ${\theta}_{\mathrm{emb}}$ govern how the angle vector $\Gamma(\mathbf{x},t)$ is generated, and their realization depends on the chosen embedding strategy.

We investigate two alternative mechanisms:

\begin{itemize}

\item {Hybrid Classical Embedding (FNN-based).}  
In this configuration, ${\theta}_{\mathrm{emb}}$ corresponds to the weights of a classical feed-forward neural network that maps the input coordinates $(\mathbf{x},t)$ to the angle vector $\Gamma(\mathbf{x},t)$. This approach enables nonlinear preprocessing in the classical domain and allows efficient computation of spatial and temporal derivatives via automatic differentiation.

\item {Fully Quantum Embedding (QNN-based).}  
In this configuration, ${\theta}_{\mathrm{emb}}$ represents the trainable parameters of a shallow auxiliary quantum circuit. Expectation values from this circuit are transformed into rotation parameters defining $U_{\text{enc}}$. Consequently, both feature generation and transformation are implemented through quantum operations, yielding a fully quantum feature map with compact parameterization.

\end{itemize}
To ensure a controlled architectural comparison, the downstream variational circuit $U_{\text{var}}({\theta}_{\mathrm{var}})$ and the physics-informed training objective remain identical across both embedding strategies. As a result, any observed differences in convergence behavior, approximation accuracy, or stability arise solely from the embedding mechanism.

\subsection{Variational Quantum Circuit Design}

Following the encoding stage, the prepared quantum state is processed by a parameterized variational quantum circuit (VQC), which forms the trainable quantum component of the proposed architecture. The purpose of this circuit is to transform the input-dependent encoded state into a latent quantum representation from which the solution field can be extracted through measurement.

The variational unitary is parameterized by a set of trainable variables ${\theta}_{\mathrm{var}}$ and is shared across all spatiotemporal collocation points. 
This shared structure reduces the number of independent quantum parameters and enforces a consistent latent representation across the domain.
The variational unitary is constructed as a layered composition of $L$ blocks,
$U_{\text{var}}({\theta}_{\mathrm{var}})
=
U_L({\theta}_L)\cdots U_2({\theta}_2)U_1({\theta}_1)$,
where each layer $U_\ell({\theta}_\ell)$ consists of parameterized single-qubit rotations followed by fixed entangling operations.
For a given input $(\mathbf{x},t)$, the full quantum state after encoding and variational processing is
\[
|\psi_{\mathrm{var}}(\mathbf{x},t)\rangle
=
U_{\text{var}}({\theta}_{\mathrm{var}})
U_{\text{enc}}(\mathbf{x},t;{\theta}_{\mathrm{emb}})
|0\rangle^{\otimes n}.
\]
The predicted solution value $\tilde{u}(\mathbf{x},t)$ is obtained by evaluating the expectation of a predefined observable $\mathcal{O}$ on this transformed state.
The variational parameters ${\theta}_{\mathrm{var}}$ are optimized jointly with the embedding parameters ${\theta}_{\mathrm{emb}}$ using the physics-informed loss defined in the subsequent subsection. By repeatedly applying a shared variational circuit to differently encoded inputs, the model captures nonlinear couplings in the solution manifold while preserving a gate structure compatible with NISQ-era hardware constraints.

\subsection{Observable and Output Mapping}
To obtain a scalar field prediction from the transformed quantum state, we evaluate the expectation value of a Hermitian observable. In this work, we define the measurement operator as
$\mathcal{O} = \bigotimes_{i=1}^{n} \sigma_z^{(i)}$,
where $\sigma_z^{(i)}$ denotes the Pauli-$Z$ operator acting on the $i$-th qubit. The predicted solution value is therefore given by

\[
\tilde{u}(\mathbf{x},t)
=
\langle \psi_{\mathrm{var}}(\mathbf{x},t) | \mathcal{O} | \psi_{\mathrm{var}}(\mathbf{x},t) \rangle.
\]
This tensor-product observable aggregates information across all qubits and maps the high-dimensional quantum state to a real-valued scalar. Since the eigenvalues of Pauli-$Z$ operators lie in $\{-1,1\}$, the resulting expectation value is naturally bounded within $[-1,1]$. This boundedness improves numerical stability during optimization and prevents uncontrolled growth of the predicted field.

The choice of $\mathcal{O}$ does not restrict generality; alternative Hermitian operators or weighted combinations of Pauli observables could be employed to modify the output scaling or capture additional structure. However, the tensor-product Pauli-$Z$ operator provides a simple, hardware-efficient measurement strategy compatible with near-term quantum devices.
For problems requiring multiple output fields, independent observables can be assigned to separate measurement channels, enabling simultaneous prediction of vector-valued quantities.

\subsection{Physics-Informed Optimization}

Model training is performed by minimizing a composite loss function constructed from residual evaluations at collocation points sampled throughout the spatiotemporal domain. The total objective is defined as
\[
\mathcal{L}({\theta}_{\mathrm{var}},{\theta}_{\mathrm{emb}}) 
=
\mathcal{L}_{\text{PDE}}
+
\lambda_{\text{BC}} \mathcal{L}_{\text{BC}}
+
\lambda_{\text{IC}} \mathcal{L}_{\text{IC}},
\]
where $\lambda_{\text{BC}}$ and $\lambda_{\text{IC}}$ are weighting coefficients that balance the contributions of boundary and initial constraints.
The individual loss components are given by

\begin{align}
\mathcal{L}_{\text{PDE}} 
&=
\sum_{(\mathbf{x}^j,t^j)\in \mathcal{S}_{\text{int}}}
\mathcal{R}_{\text{PDE}}(\mathbf{x}^j,t^j)^2, \\
\mathcal{L}_{\text{BC}} 
&=
\sum_{(\mathbf{x}^j,t^j)\in \mathcal{S}_{\text{bd}}}
\mathcal{R}_{\text{BC}}(\mathbf{x}^j,t^j)^2, \\
\mathcal{L}_{\text{IC}} 
&=
\sum_{\mathbf{x}^j\in \mathcal{S}_{\text{init}}}
\mathcal{R}_{\text{IC}}(\mathbf{x}^j)^2.
\end{align}

The collocation sets $\mathcal{S}_{\text{int}}$, $\mathcal{S}_{\text{bd}}$, and $\mathcal{S}_{\text{init}}$ correspond to interior, boundary, and initial sampling points, respectively.
Spatial and temporal derivatives appearing in $\mathcal{R}_{\text{PDE}}$ are computed through automatic differentiation applied to the hybrid computational graph. Gradients with respect to the variational quantum parameters ${\theta}_{\mathrm{var}}$ are evaluated using parameter-shift differentiation, while gradients with respect to the embedding parameters ${\theta}_{\mathrm{emb}}$ are computed either classically (for the FNN-based embedding) or via quantum gradient rules (for the QNN-based embedding).

The complete parameter set 
$\Theta = ({\theta}_{\mathrm{var}}, {\theta}_{\mathrm{emb}})$
is optimized jointly using gradient-based methods. By enforcing the governing equation and constraints directly in the objective function, the model learns a solution that satisfies the PDE without requiring labeled solution data.

\section{Numerical Results}
In this section, we discuss the experimental results of the quantum-assisted learning model applied to the Heat equation, one of the most significant parabolic PDEs in physics. We use a classical PINN architecture with 4 hidden layers of 50 neurons each as the baseline for comparison. Regarding the quantum-assisted learning models, we employ two architectural variants of the quantum PINN that differ only in their embedding components. In the first architecture, the embedding is a feed-forward neural network with 2 hidden layers and 10 neurons per layer. The second architecture utilizes a quantum neural network for embedding, with the number of qubits and layers matching those of the primary trainable quantum circuit. For the primary trainable quantum component, both architectures share a standard VQC based on a hardware-efficient ansatz (HEA). The specific parameters of the variational quantum circuit will be described in detail in the following sections for one- and two-dimensional cases.

\begin{figure}[t]
    \centering
    \includegraphics[scale=0.17]{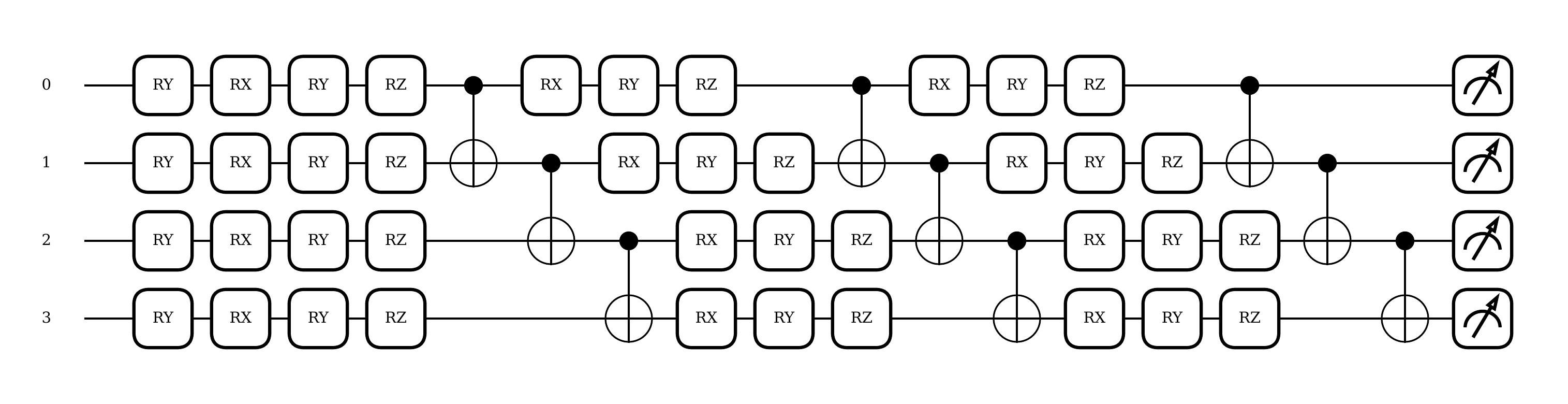}
    \caption{
    \small Variational quantum circuit of the Heat Equation.}
    \label{fig:heat_1d_circuit}
    \vspace{-0.5cm}
\end{figure}

\begin{figure}[t]
    \centering
    \includegraphics[scale=0.22]{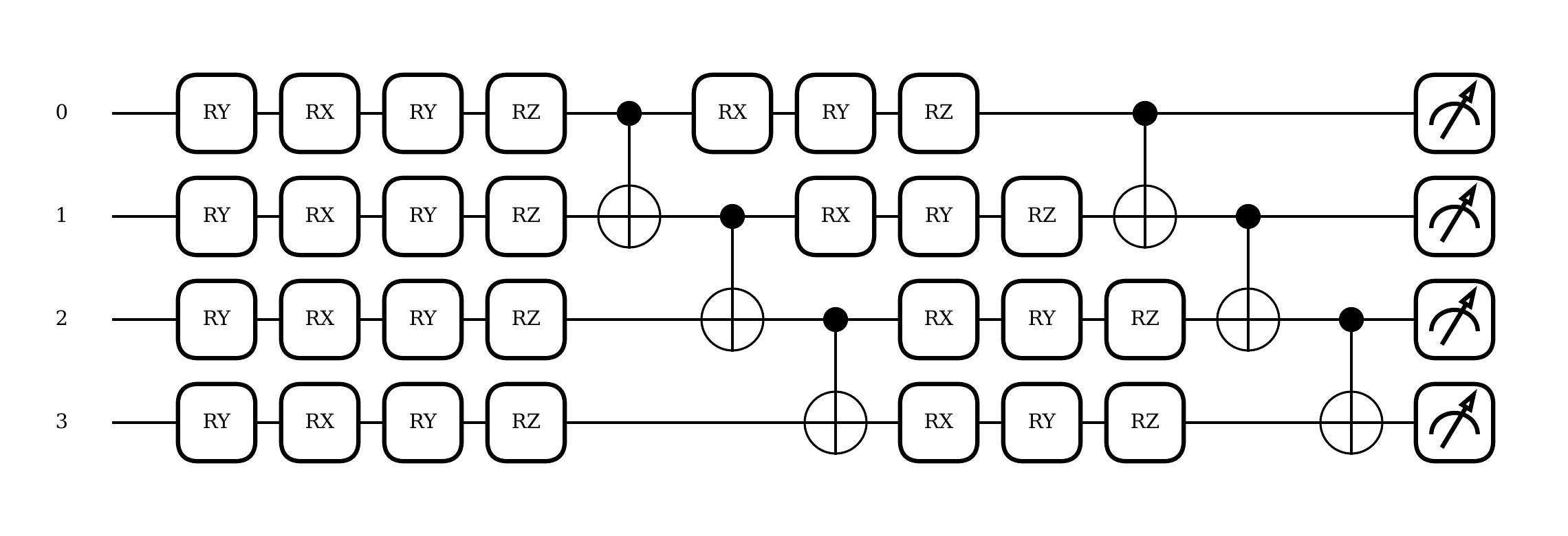}
    \caption{
    \small Embedding quantum circuit for the Heat equation.}
    \label{fig:heat_2d_circuit}
    \vspace{-0.5cm}
\end{figure}

\subsection{Enviroment Setup}

\begin{figure}[t]
    \centering
    \includegraphics[scale=0.35]{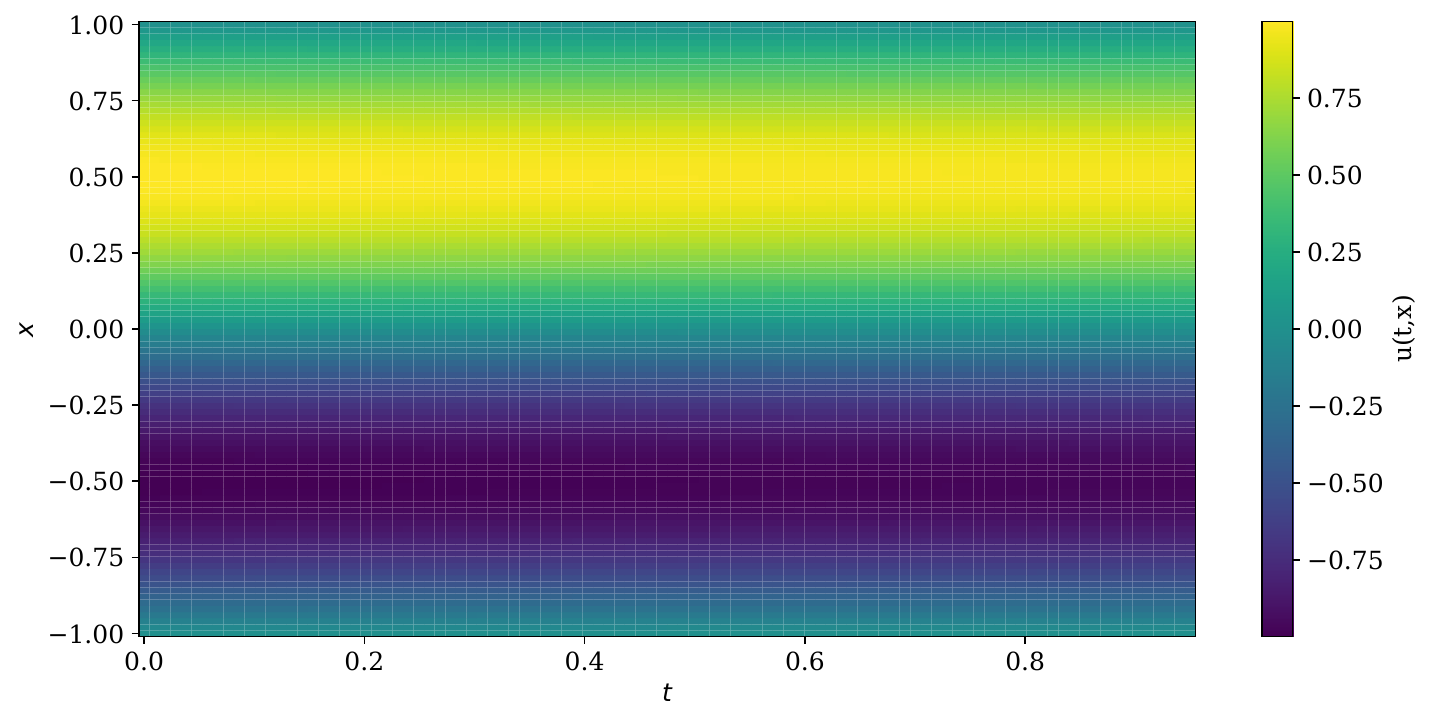}
    \caption{
    \small Exact solution solved by the RK45 method for the one-dimensional Heat equation..}
    \label{fig:heat_1d_reference}
    \vspace{-0.5cm}
\end{figure}

\begin{figure}[htbp]
    \centering
    \includegraphics[scale=0.25]{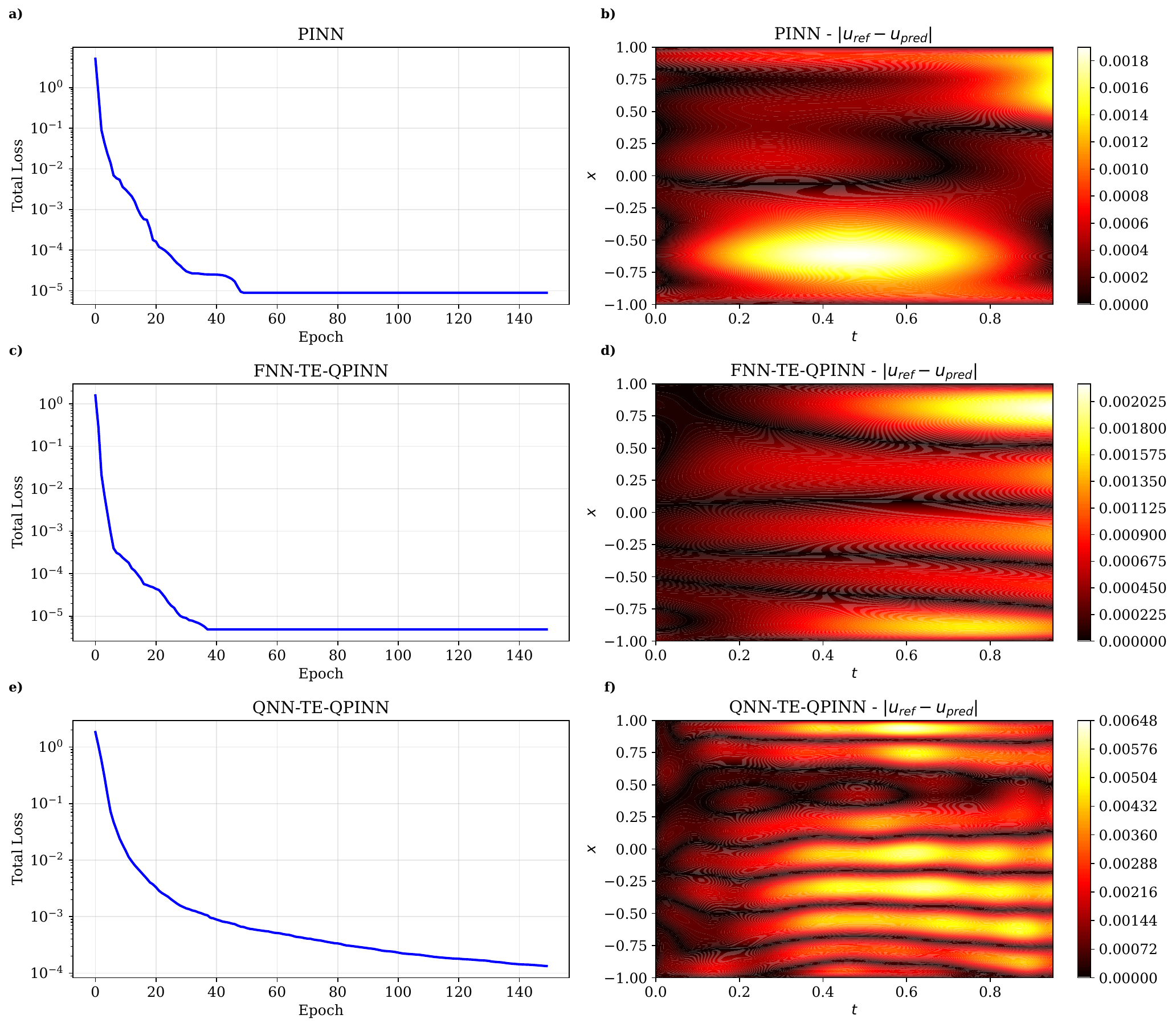}
    \caption{
    \small Performance comparison results among PINN, FNN-TE-QPINN, and QNN-TE-QPINN Models for one-dimensional Heat equation after 150 training epochs. (a) \& (b) The loss function of the PINN model and the absolute error between its solution and the RK45 reference solution. (c) \& (d) The loss function and absolute error of the FNN-TE-QPINN model. (e) \& (f) The loss function and absolute error of the QNN-TE-QPINN model.}
    \label{fig:heat_1d_comparison}
    \vspace{-0.5cm}
\end{figure}

\begin{figure*}[htbp]
    \centering
    \includegraphics[scale=0.4]{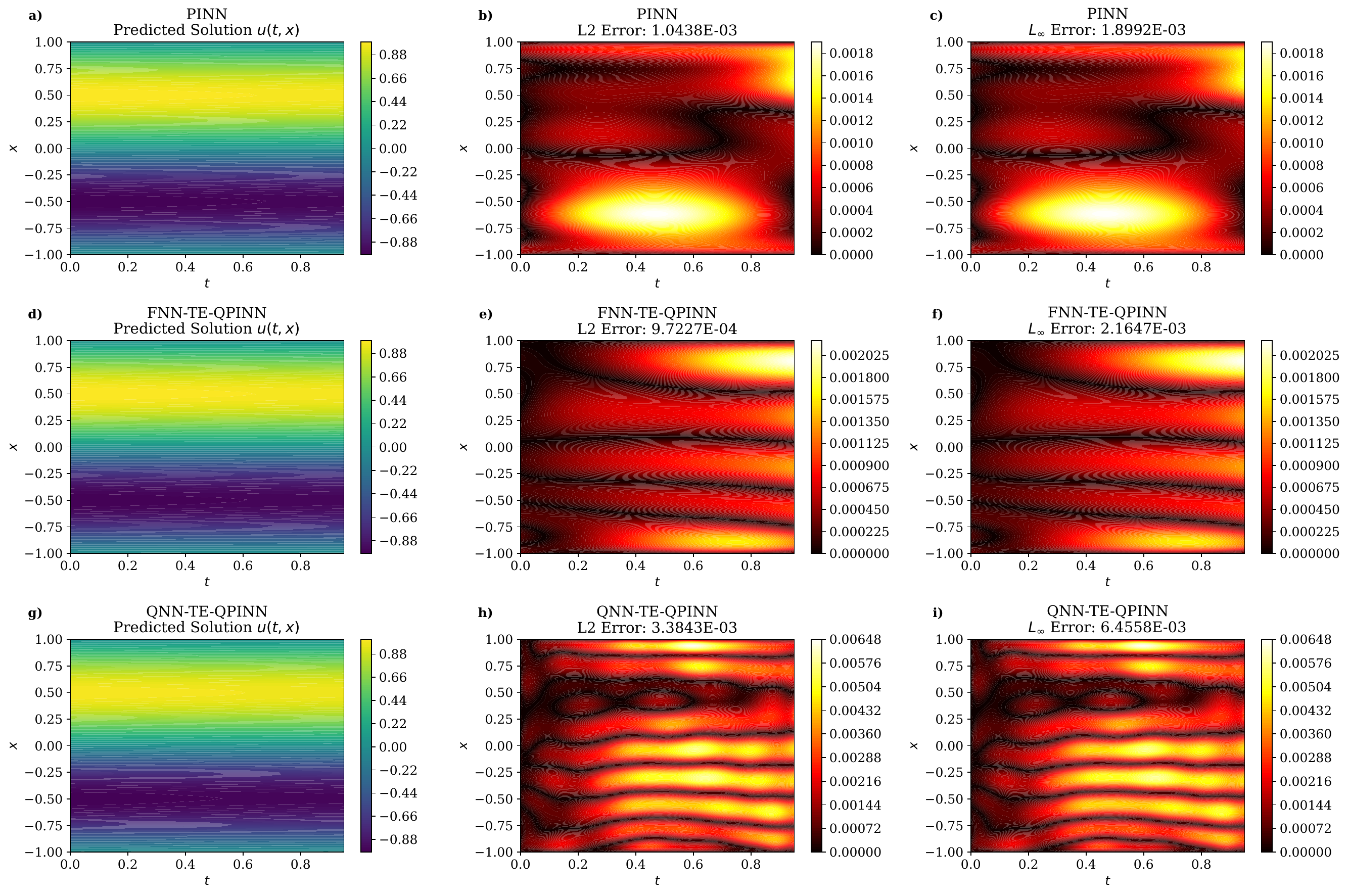}
    \caption{
    \small Inference results of trained PINN, FNN-TE-QPINN, and QNN-TE-QPINN Models for the one-dimensional Heat equation. (a), (b), and (c) The predicted solution, L2 relative error, and L-infinity relative error of the PINN model. (d), (e), and (f) The predicted solution, L2 relative error, and L-infinity relative error of the FNN-TE-QPINN model. (g), (h), and (i) The predicted solution, L2 relative error, and L-infinity relative error of the QNN-TE-QPINN model.}
    \label{fig:heat_1d_inference}
    \vspace{-0.5cm}
\end{figure*}

\begin{figure*}[htbp]
    \centering
    \includegraphics[scale=0.35]{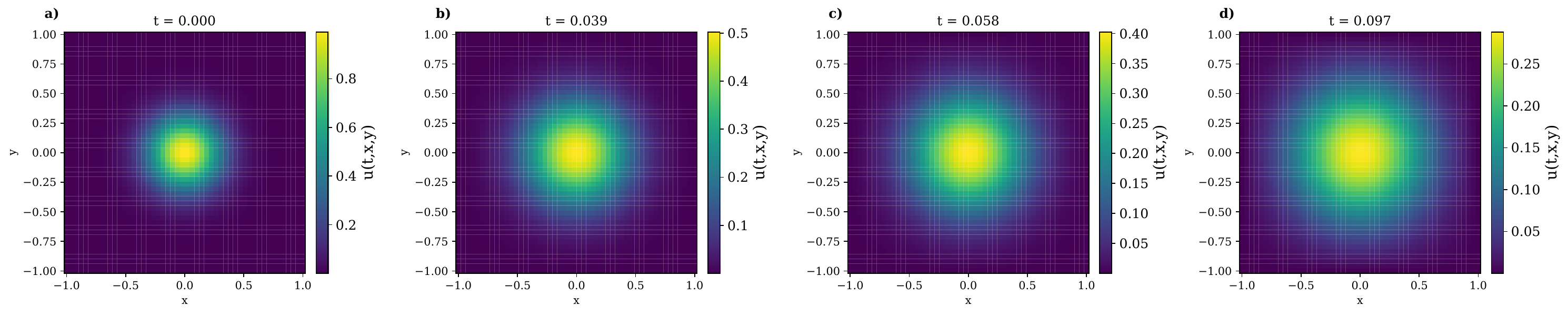}
    \caption{
    \small Exact solution solved by the RK45 method for the two-dimensional Heat equation at time $t = 0$, $t = 0.039$, $t = 0.058$, and $t = 0.097$.}
    \label{fig:heat_2d_reference}
    \vspace{-0.5cm}
\end{figure*}

For the software implementation, Python was chosen as the primary programming language. The deep learning library PyTorch \cite{pytorch}
was employed to construct classical neural networks and optimize loss functions for both classical tensors and quantum circuits. For the trainable quantum components, we utilized the PennyLane \cite{pennylane} quantum machine learning library by Xanadu to build the circuits, perform encoding, and execute measurements to collapse the quantum states into classical bitstrings. In the classical feed-forward neural network used for embedding, we implemented the tanh activation function to ensure the output remains compatible with the input rotation angles $\theta$, which range from $- \pi$ to $ \pi$. For the primary quantum learning circuit, we employed the L-BFGS optimizer—a Quasi-Newton method available in PyTorch—to minimize the objective function. 
Within the L-BFGS configuration, we applied the strong\_wolfe line search condition to ensure optimal step sizes during optimization. Regarding the hardware for our experiments, we utilized a GPU compute node provided by the REmotely-managed Power-Aware Computing Systems and Services (REPACSS) \cite{repacss}. The node is equipped with dual Intel Xeon Gold 6448Y processors totaling 64 cores and 512 GB of RAM. Each GPU compute node features four NVIDIA H100 NVL GPUs, each with 94 GB of VRAM and 1.92 TB of local storage.

\subsection{Results and Analysis}

\begin{figure}[htbp]
    \centering
    \includegraphics[scale=0.25]{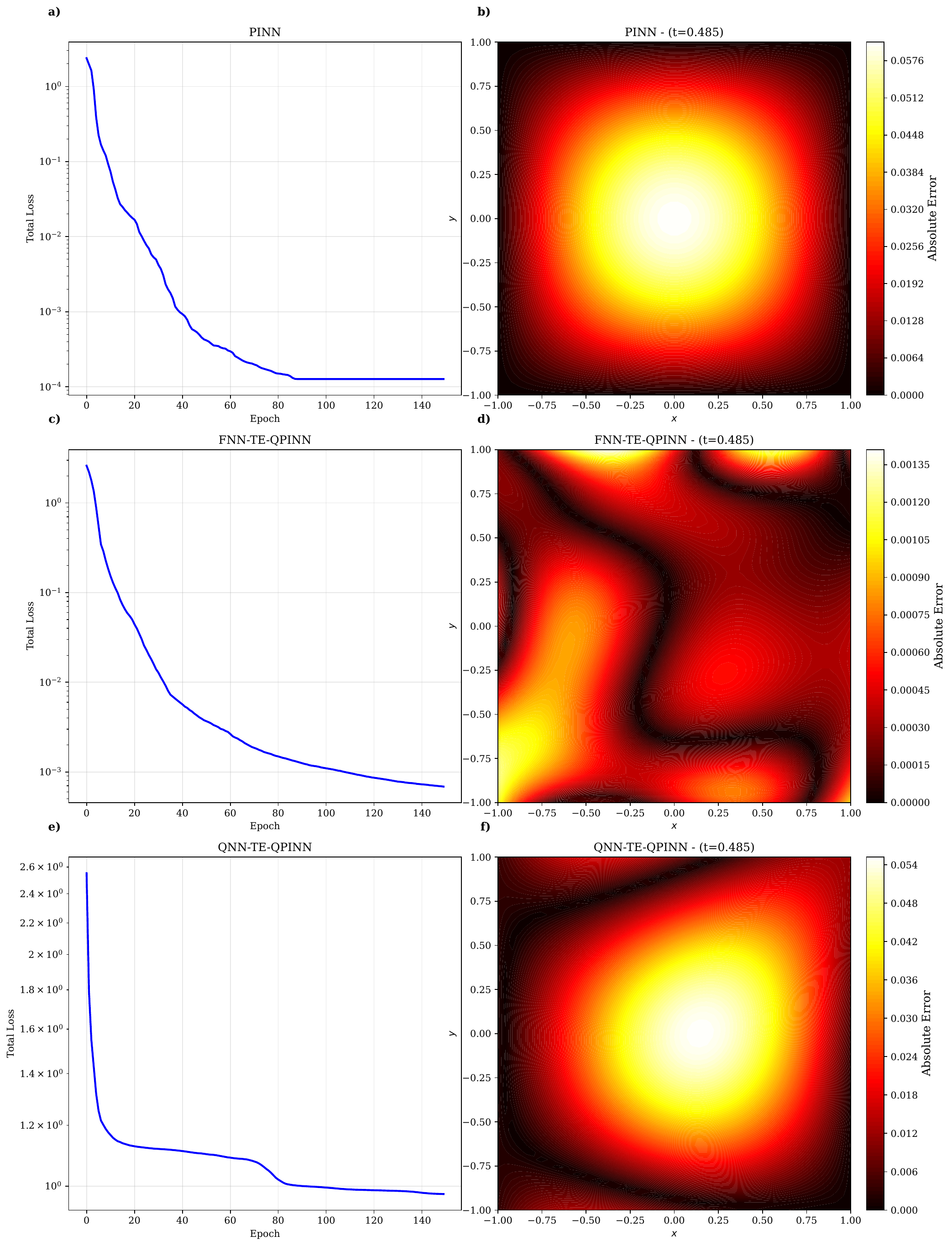}
    \caption{
    \small Performance comparison results of the PINN, FNN-TE-QPINN, and QNN-TE-QPINN models after 150 training epochs. (a) \& (b) The evolution of the loss function during training and the absolute error of the PINN model at time t = 0.485. (c) \& (d) Results for the FNN-TE-QPINN model. (e) \& (f) Results for the QNN-TE-QPINN model.}
    \label{fig:heat_2d_comparison}
    \vspace{-0.5cm}
\end{figure}

The detailed experiments for the Heat equation are divided into two scenarios: one-dimensional (1D) and two-dimensional (2D). Based on the hardware configuration described in Section III-A, we utilized 6 qubits for the 1D case (Fig. \ref{fig:heat_1d_circuit}) and 4 qubits for the 2D case (Fig. \ref{fig:heat_2d_circuit}). Our experiments indicate that using at least 2 qubits and 3 layers allows the FNN-TE-QPINN model to outperform the classical PINN in terms of solution quality. Leveraging the GPU configuration provided by REPACSS, we trained models with up to 8 qubits and 20 layers for the 1D case, and 6 qubits with 15 layers for the 2D case. While the number of qubits and layers could be further scaled using parallel processing methods such as MPI, we found this to be unnecessary for the scope of the current study. These values represent the maximum parameters that can be simulated on classical hardware, given our current configuration. If necessary, these parameters could be further increased by using Message Passing Interface (MPI) for parallel programming.

\subsubsection{One-dimensional Heat equation}

\begin{figure*}[htbp]
    \centering
    \includegraphics[scale=0.4]{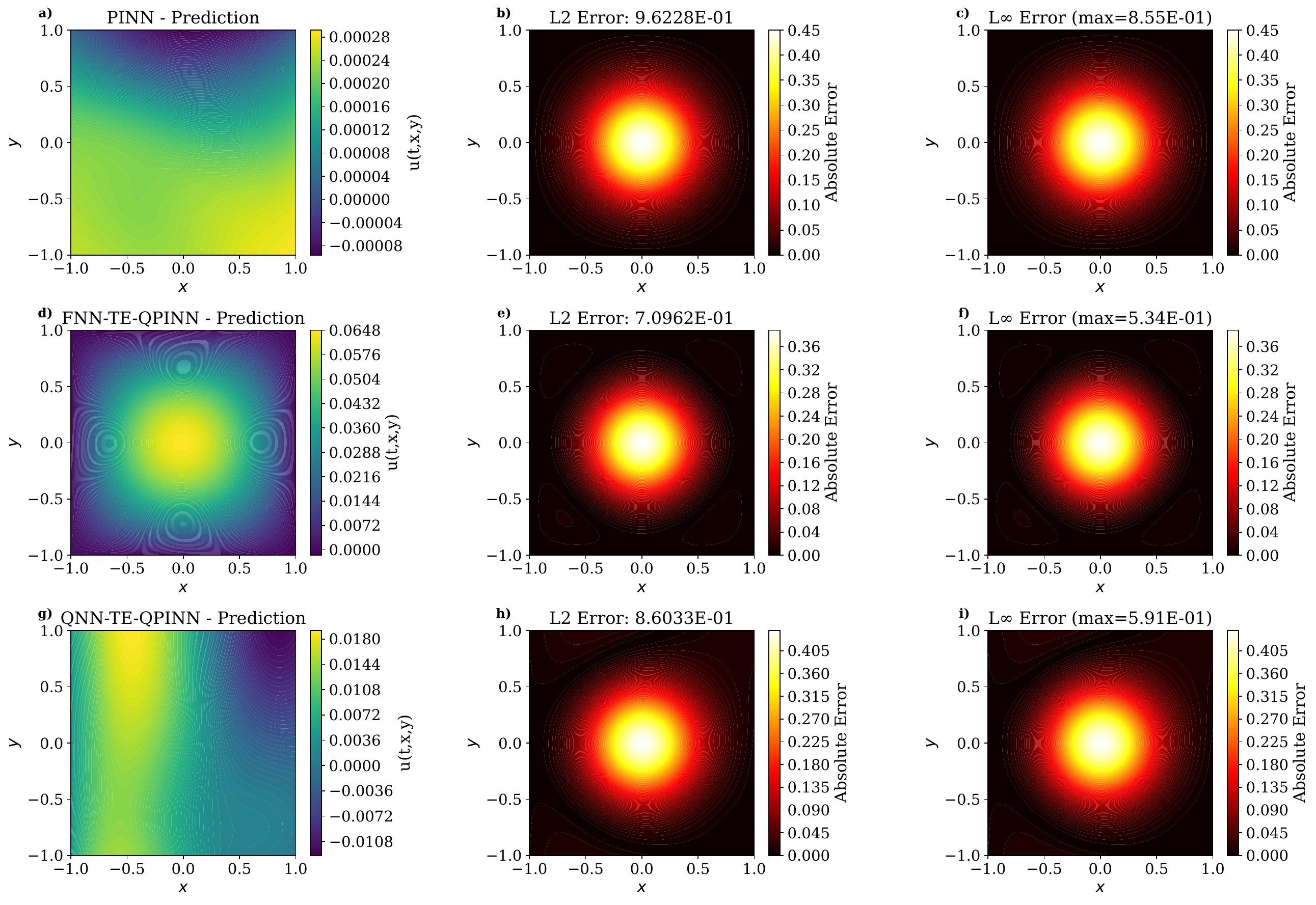}
    \caption{
    \small Inference results of trained PINN, FNN-TE-QPINN, and QNN-TE-QPINN Models for the two-dimensional Heat equation.  (a), (b), and (c) The predicted solution, L2 relative error, and L-infinity relative error of the PINN model. (d), (e), and (f) The predicted solution, L2 relative error, and L-infinity relative error of the FNN-TE-QPINN model. (g), (h), and (i) The predicted solution, L2 relative error, and L-infinity relative error of the QNN-TE-QPINN model.}
    \label{fig:heat_2d_inference}
    \vspace{-0.5cm}
\end{figure*}

To evaluate the performance of both classical and quantum machine learning methods, we use the Runge-Kutta 45 (RK45) method to compute 
a reference solution of the differential equation. Fig. \ref{fig:heat_1d_reference} illustrates the solution for the one-dimensional Heat equation presented as a heatmap, with the time axis $t$ ranging from 0 to 1 and the spatial axis $x$ ranging from -1 to 1. In the 1D scenario, the diffusion coefficient is set to a relatively small value of $0.01/\pi$, which corresponds to materials with low thermal conductivity.

Fig. \ref{fig:heat_1d_comparison} compares the performance of the three models—PINN, FNN-TE-QPINN, and QNN-TE-QPINN after training 150 epochs—and evaluates their solutions against the exact reference solution obtained via the RK45 method. The results demonstrate that the FNN-TE-QPINN model significantly outperforms both the classical PINN and the QNN-TE-QPINN. The hybrid approach of combining a classical neural network for embedding with a quantum circuit yielded high precision, with the FNN-TE-QPINN achieving a total loss of $4.8670 \times 10^{-6}$, while the PINN and QNN-TE-QPINN reached $8.9515 \times 10^{-6}$ and $1.3406 \times 10^{-4}$, respectively. From the loss function plots, it is evident that FNN-TE-QPINN converges at 40 epochs, whereas the classical PINN requires nearly 50 epochs. Both curves exhibit signs of oscillations (sawtooth patterns). Although QNN-TE-QPINN has the highest loss among the three models, its curve shows a smooth trajectory, indicating effective learning without signs of overfitting. This smooth represents a significant advantage for this model with its purely quantum embedding component. In terms of accuracy, the FNN-TE-QPINN exhibited a remarkably low maximum absolute error of $4.5431 \times  10^{-7}$, which is better than the PINN at $5.2349 \times  10^{-7}$ and the QNN-TE-QPINN at $5.5036 \times  10^{-6}$. Although the QNN-TE-QPINN—a fully quantum architecture for both embedding and the primary learning circuit—is often highly anticipated, it produced the most significant error in this specific case. This detriment may be attributed to the fact that the quantum embedding parameters were not as optimized as those of the FNN. Interestingly, despite the higher numerical error, the heatmap for the QNN-TE-QPINN appears visually "cleaner" and matches the exact solution's color profile more closely than the other two methods.

Fig. \ref{fig:heat_1d_inference} illustrates the inference results of the three models. It is observed that all three models produce solutions that are in close agreement with the reference ground truth. Among them, the FNN-TE-QPINN achieves the highest accuracy, yielding an $L_{2}$ absolute error of $9.7227 \times 10^{-4}$ and an $L_{\infty}$ ($L_{max}$) absolute error of $2.1647 \times 10^{-3}$. The classical PINN ranks second, with respective values of $1.0438 \times 10^{-3}$ and $1.8992 \times 10^{-3}$. The QNN-TE-QPINN ranks third, with an $L^2$ error of $3.3843 \times 10^{-3}$ and an $L_{\infty}$ error of $6.4558 \times 10^{-3}$. These results demonstrate that combining classical neural networks with quantum circuits yields exceptional performance in learning the Heat equation, with an average error of nearly 0.1\%. Although the QNN-TE-QPINN model achieved a slightly higher average error of 0.3\%, this is considered a very positive outcome for an architecture utilizing a purely quantum circuit for embedding.

\subsubsection{Two-dimensional Heat equation}

In this section, we describe the experimental setup for solving the Heat equation in two-dimensional (2D) space. We utilize a parameterized quantum circuit consisting of 4 qubits and 10 layers, as illustrated in Fig. \ref{fig:heat_2d_circuit}. The classical exact solution used as a reference was computed using the RK45 method and can be seen in Fig. \ref{fig:heat_2d_reference} at specific time intervals: $t=0$, $t=0.039$, $t=0.058$, and $t=0.097$. To accommodate the hardware constraints and prevent memory overflow, we reduced the input spatial grid size by a factor of four, resulting in 50 points per dimension and the time domain is set to 0 to 0.1, rather than 0 to 1 as in the one-dimensional case. In the two-dimensional case, the diffusion coefficient of the equation is set to a relatively high value of $2/\pi$ to align with the short time horizon $t$. Physically, this coefficient corresponds to highly thermally superconductive materials. While such a large coefficient poses a significant challenge for traditional exact numerical methods like RK45 (due to the constraints of finite-difference schemes), it offers a primary advantage for approximation-based methods such as PINN and QPINN. The input dataset for the 2D models consists of 2,500 initial-condition points, 9,604 boundary-condition points, and 112,896 interior points. In total, 125,000 collocation points are utilized for training each model. Regarding model complexity, the classical PINN has 271 parameters, while the FNN-TE-QPINN, which combines classical and quantum components, has 8,114 parameters. In contrast, the QNN-TE-QPINN has the fewest parameters, with only 96.

Fig. \ref{fig:heat_2d_comparison} presents the performance results of the three models after training for 150 epochs. These results are highly consistent with the performance trends observed in the one-dimensional case for all three models. The FNN-TE-QPINN demonstrates the most optimal performance, achieving a loss function value of $1.2013 \times 10^{-4}$. In comparison, the classical PINN reaches a loss of $3.9738 \times 10^{-3}$, while the QNN-TE-QPINN exhibits a significantly higher loss function value of $1.6305$. Furthermore, the right-hand column of the figure shows the absolute errors for the three models at $t = 0.485$. The FNN-TE-QPINN yields an exceptionally low maximum absolute error of only 0.00135. In contrast, the maximum errors for the PINN and QNN-TE-QPINN models are 0.0576 and 0.054, respectively.

The accuracy of the three models in solving the equation is further illustrated in the inference section of Fig. \ref{fig:heat_2d_inference}. The predictions from the FNN-TE-QPINN show the closest agreement with the reference solution, achieving an $L_2$ absolute error of $7.0962 \times 10^{-1}$ and an $L_{\infty}$ ($L_{max}$) absolute error of $5.34 \times 10^{-1}$. The classical PINN produces more modest results, with values of $9.6228 \times 10^{-1}$ and $8.55 \times 10^{-1}$, respectively. The QNN-TE-QPINN performs less effectively compared to the other two models, with an $L_2$ error reaching $8.6033 \times 10^{-1}$ and an $L_{\infty}$ error of $5.91 \times 10^{-1}$.

In summary, both PINN and QPINN models provide high-quality solutions to the Heat equation. The FNN-TE-QPINN, featuring an embedding component trained with a neural network, yields the best results. These better results underscore that enhancing the embedding process significantly empowers quantum models to achieve stronger learning capabilities and outperform classical counterparts. However, adopting a purely quantum approach for the entire embedding stage, as in QNN-TE-QPINN, remains less feasible at present, as its performance does not yet surpass that of the classical PINN.

\section{Conclusions}

Solving partial differential equations using machine learning methods has demonstrated significant potential and remains one of the most compelling research directions in the field. In this study, we leveraged the architectural advantages of quantum machine learning, combined with the proven robustness of classical neural networks, to solve both one- and two-dimensional Heat equations. Our results indicate that hybrid quantum-classical models achieve high accuracy in solving the Heat equation. Furthermore, this research paves the way for practical applications of these models in solving other types of complex nonlinear partial differential equations in the future.

However, this research also highlights that implementing a purely quantum architecture—where both the embedding and the primary learning circuit rely solely on quantum circuits—does not yet yield optimal results. This result underscores the critical role of classical models in scenarios where data processing still requires efficient transitions between classical data and quantum states. As quantum computing remains in the noisy intermediate-scale quantum era, where noise continues to degrade quantum properties, this interface remains a significant challenge that researchers must continue to address.


Our future research will continue to focus on addressing the embedding problem using both classical and quantum machine learning methods. The application of classical machine learning and deep learning architectures to the optimization of quantum systems is a promising research direction that could yield breakthrough results in the near future. We hope that this study contributes a foundational step toward this exciting and high-potential field.

\bibliographystyle{IEEEtran}
\bibliography{heat}

\end{document}